\documentclass[journal=jacsat,manuscript=article, amsmath,amssymb,]{achemso}
\usepackage{mathtools, nccmath}
\usepackage{amssymb}
\usepackage{enumerate}
\usepackage{hyperref}

\usepackage[version=3]{mhchem} 
\usepackage{color,soul}

\allowdisplaybreaks 
\author{Cian C. Reeves}
\affiliation{%
Department of Physics, University of California, Santa Barbara, Santa Barbara, CA 93117
}%
\author{Michael Kurniawan}
\affiliation{%
Materials Department, University of California, Santa Barbara, Santa Barbara, CA 93117
}
\author{Yuanran Zhu}
\affiliation{Applied Mathematics and Computational Research Division, Lawrence Berkeley National Laboratory,Berkeley, CA 94720, USA}
\author{Nikil Jampana}
\affiliation{%
Department of Physics, University of California, Santa Barbara, Santa Barbara, CA 93117
}%
\author{Jacob Brown}
\affiliation{%
Department of Chemistry and Biochemistry, University of California, Santa Barbara, Santa Barbara, CA 93117
}%
\author{Chao Yang}
\affiliation{Applied Mathematics and Computational Research Division, Lawrence Berkeley National Laboratory,Berkeley, CA 94720, USA}
\author{Khaled Z. Ibrahim}%
\affiliation{Applied Mathematics and Computational Research Division, Lawrence Berkeley National Laboratory,Berkeley, CA 94720, USA}
\author{Vojtech Vlcek}
\affiliation{%
Department of Chemistry and Biochemistry, University of California, Santa Barbara, Santa Barbara, CA 93117
}%
\altaffiliation{%
Materials Department, University of California, Santa Barbara, Santa Barbara, CA 93117
}

\title{A Practical Framework for Simulating Time-Resolved Spectroscopy Based on a Real-time Dyson Expansion}

\begin{document}
\begin{abstract}
  Time-resolved spectroscopy is a powerful tool for probing electron dynamics in molecules and solids, revealing transient phenomena on sub-femtosecond timescales. The interpretation of experimental results is often enhanced by parallel numerical studies, which can provide insight and validation for experimental hypotheses. However, developing a theoretical framework for simulating time-resolved spectra remains a significant challenge. The most suitable approach involves the many-body non-equilibrium Green’s function formalism, which accounts for crucial dynamical many-body correlations during time evolution. While these dynamical correlations are essential for observing emergent behavior in time-resolved spectra, they also render the formalism prohibitively expensive for large-scale simulations. Substantial effort has been devoted to reducing this computational cost---through approximations and numerical techniques---while preserving the key dynamical correlations. The ultimate goal is to enable first-principles simulations of time-dependent systems ranging from small molecules to large, periodic, multidimensional solids.  In this perspective, we outline key challenges in developing practical simulations for time-resolved spectroscopy, with a particular focus on Green’s function methodologies. We highlight a recent advancement toward a scalable framework: the real-time Dyson expansion (RT-DE) \cite{Reeves_2024}. We introduce the theoretical foundation of RT-DE and discuss strategies for improving scalability, which have already enabled simulations of system sizes beyond the reach of previous fully dynamical approaches. We conclude with an outlook on future directions for extending RT-DE to first-principles studies of dynamically correlated, non-equilibrium systems.
\end{abstract}
\section{Introduction }

Understanding the non-equilibrium properties of molecules and solids is important for many areas of chemistry, physics and materials science.  For example, charge transport properties of electrons in organic semiconductors and molecular crystals, perovskites, or transition metal dichalcogenides are key for designing improved photovoltaics, light emitting diodes and new organic transistors\cite{Ruiz_2012,Brenner2016,Schmidt_2015,sangwan_2018,Huo_2015,May_2012,ovchinnikov_2014,Wu2020,Groves_2017,Lee_2018,Zhou_2018,Troisi_2011,Tang2009,Fratini_2016}.  Another expanding area in non-equilibrium science is studying the dynamics of emergent quantum states and quasiparticles such as the transport and decoherence properties of electron spins \cite{Beaurepaire_1996,giovanni_2015,ping_2018,Xu2020,Xu2024,Park_2020,Park_2022,Park_2022_2}, critical for spintronics and in designing quantum computing platforms, and the dynamical properties of excitons\cite{Ryasnyanskiy_2011,wolf_2021,dimitriev_2022,Ovesen2019,Perea-Causin_2022,Li_2025,chen_2020,chen_2022,zhang_2011,chan2024}, important in the optoelectronic properties of low-dimensional materials with strong exciton binding energies for light harvesting and quantum sensing applications.  Besides non-equilibrium physics being important for understanding a material's behavior, in recent years there has been large interest in leveraging these transient regimes in new devices with properties that can be switched by coupling to external fields. This includes Floquet engineering, where a materials properties can be tuned through continuous periodic driving\cite{Nuske_2020,Zhou_2023,wang_2013,takashi_2019,cheng_2019}, as well as phenomena occurring on ultrafast timescales such as petahertz switching\cite{Ossiander2022,Kawakami2020,Mashiko2018,Garg2016}, ultrafast phase transitions\cite{Disa_2023,Beebe_2017,Bao_2022,Hervé2024,Johnson2023,dringoli_2024,park_2024} and the creation and dynamics of exotic states of matter\cite{Bao_2022,Dong_2021,Hedayat_2021,Siwick2021}.  

The field of ultrasfast physics in particular is attracting great attention in recent years due to: 1) advances in pump-probe spectroscopy techniques such as time-resolved angle resolved photoemission spectroscopy (TR-ARPES) and transient extreme ultraviolet spectroscopy (T-XUVS) that allow systems to be probed on femto and even attosecond timescales\cite{sobota_ultrafast_2012,bovensiepen_elementary_2012,Schmitt_ultrafast_2011,bressler_ultrafast_2004,lin_carrier_2017,Liu_2023,cushing_differentiating_2019,Dong_2021,perfetti_2006,Boschini_timeresolved_2024,Rivas_2022,Quintero_2025,Biswas_2024}, and 2) promises of devices leveraging ultrafast phenomena  to push the boundaries of metrology, optoelectronics and quantum computing\cite{Mashiko2018,Garg2016,Schiffrin2013,Schultze2013,Krausz2014,Zong2023,clark_2007,Wang2017}.  

While ultrafast spectroscopy provides a unique window into non-equilibrium systems, the interpretation and explanation of experimental observations is often a difficult task.  Factors like noise and impurities present in real experiments are compounded by the inherent complexity of a time-evolving system.  Thus in practice it can be difficult to unambiguously determine the mechanisms behind observed phenomena.  Theory is often used alongside experiment as a secondary source of confirmation and can offer more flexibility, aiding in verification of proposed hypotheses.  For example, by simulating the same system with different parts of the model switched on or off (e.g, thermal effects, electron-phonon interactions, spin-orbit coupling, etc.) one can determine more concretely what contributes to specific observations.  Therefore, in parallel to advances in time-resolved experimental techniques there has been a broad effort in the theory community to develop practical computational tools to study time-dependent systems and to work in conjunction with experiment to improve understanding of non-equilibrium phenomena. 

Developing efficient non-equilibrium simulations is challenging as it takes the already difficult problem of simulating equilibrium quantum systems and adds the additional complexities of coupling to time-dependent fields, tracking the systems evolution and---depending on observables of interest---accurately capturing dissipation effects.  Exact solutions to this problem scale exponentially with system size making them intractable for anything beyond small systems. In practice approximations are necessary, requiring trade-offs between accuracy and computational cost.  Different approaches are employed depending on a variety of factors, such as the quantities of interest (e.g., total energies, photoemission spectra, or full wavefunctions), the type of time evolution (e.g., steady-state, Floquet, or ultrafast dynamics), or the systems of interest (e.g., molecular, solid-state, or lattice model systems).

At one extreme are methods that are based on simulating the time-evolution of the many-body wavefunction.   These consist of approaches such as time-dependent full configuration interaction\cite{Innerberger2020,Reeves_2025,peng_2018}, which is equivalent to solving the full Schr\"odinger equation as well as truncated versions of this like time-dependent multi-configurational Hartree-Fock (TD-MCHF)\cite{dalgaard_1980,lode_2020,Hochstuhl_2011} or time-dependent (reduced) active space CI (TD-(R)ASCI)\cite{hochstuhl_2012,sato_2013,bauch_2014,Miyagi_2013}.  These approaches are highly accurate and give access to the full wavefunction and thus all the properties of the system, however they become exponentially more difficult to solve as system size increases making them impractical outside of small molecular and atomic systems.  Non-exponentially scaling wavefunction methods  can be used to go beyond CI type methods however they are still limited. For example time-dependent coupled cluster (TD-CC) is used in quantum chemistry to simulate small weakly correlated molecular and atomic systems\cite{skeidsvoll_comparing_2023,skeidsvoll_time-dependent_2020,walz_application_2012,koulias_relativistic_2019,sato_communication_2018,pathak_2020,pathak_2021,pathak_time-dependent_2020} and time-dependent density matrix renormalization group (TD-DMRG) which can be applied in large scale systems but only those with suitable entanglement entropy structure \cite{cazalilla_2002,schollwock_2006,Daley_2004,zwolak_2004,schollwock_2005,schollwock_2011}.    

On the opposite end of the complexity spectrum, significant progress can be made towards simulating large scale systems from first principles by compressing the information: the observables of interest are not computed from the wavefunction but rather from functionals of the one particle density, as in the case of time-dependent density functional theory (TD-DFT)\cite{runge_1984,ullrich2012time,marques2006time}, or the one particle density matrix (1-RDM) typically used in quantum master equation and Lindblad type dynamics.  These approaches are routinely used to simulate quantum transport\cite{Stefanucci_2004,Vyas_2020,zheng_2010,li_2005,fischetti_1999,Fischetti_1998,harbola_2006,xu_2021,xu_2024_2,Xu2020,Xu2024,Zhanghui_2019,krieger_2015}, excited state properties\cite{Hehn_2022,HERBERT2023,Jin_2023,Jacquemin_2011,Furche_2002} and even ultrafast non-equilibrium dynamics\cite{turkowski_2008,Tancogne-Dejean_2018,oliveira_2015,Lian_2018,zhang_2023,mengxue_2022,Beaulieu_2021}.  An added complexity arises when applying these methods to model time-resolved spectroscopy, due to the limited information contained in the single particle density and single particle density matrix. Ultimately it requires the simulation of the full time-resolved spectroscopy experiment in order to extract the transient spectra\cite{degiovannini_2017,degiovannini_2020,DEGIOVANNINI2022}.  A further downside of TD-DFT is that, while in principle exact, it relies on knowledge of a fictitious potential known as the exchange-correlation potential.  This turns out to be a major issue since its exact form is not known and it is difficult to make systematically improved approximations to it.  In principle this potential is non-local in time, which introduces important memory effects into the time-evolution, however the majority of TD-DFT studies use time local approximations and neglect memory effects entirely.  

Falling somewhere between TD-DFT and wavefunction based methods in terms of computational cost and accuracy is the many-body Green’s function (GF) approach based on many-body perturbation theory (MBPT). Arguably a gold standard for equilibrium materials simulations\cite{golze_2019} it is directly related to the ARPES/TR-ARPES spectrum\cite{martin2016interacting,freericks_2009,stefanucci2013nonequilibrium} therefore offering a more natural route to it's simulation from first principles compared to TD-DFT.  The TD-DFT and GF methodologies share similarities in that both downfold the full many-body problem onto a single particle picture and both depend on the knowledge of a single space-time non-local potential---referred to as the self-energy in the Green's function formalism---. Furthermore, as with TD-DFT's---at least in theory---exchange correlation potential,  the self-energy introduces dependence on the system's past into the time-evolution. A major advantage the GF approach holds over TD-DFT is that although the exact form of the self-energy is not known, outside of small, exactly solvable systems, there is a well defined procedure to approximate it and to systematically improve upon these approximations\cite{Hedin_1965,martin2016interacting,Mejuto_2022,Stefanucci_2014,Kadanoff_1962}. Unfortunately, despite the non-equilibrium GF (NEGF) formalism being known for six decades\cite{Keldysh_1964} and the advantages it holds, it has not found wide adoption beyond relatively simple model problems.  This challenge arises primarily from the structure of the Green's function equations of motion, known as the Kadanoff–Baym equations (KBEs). Solving the KBEs scales cubically with the duration of the time evolution, due to the memory effects introduced by the time-nonlocal self-energy, which manifest in the equations of motion as integrals over the systems past states.  Several attempts have been made to reduce the scaling of the GF equations of motion, however until recently these approaches have either focused on the time evolution of the equal time GF (equivalent to the density matrix and not applicable to TR-ARPES) or have failed to reduce the computational cost to the ultimately desired linear scaling needed for practical ab-initio simulations\cite{Lipavsky_1986,kaye_2021,schlunzen_2020,Perfetto_2022,chan_2021}.

Seeing a clear need for a practical framework for the efficient simulation of time-resolved spectral properties we have recently introduced a method to addresses this\cite{Reeves_2024}. Our method, the real-time Dyson expansion (RT-DE), allows for the selective reconstruction of the non-equal time GF, the key ingredient in time-resolved spectra, while also including dynamical correlation effects, crucial for properly accounting for the many-body nature of the system, without explicit integral evaluation.  This allows for the GF to be evolved only within the window which is probed by the TR-ARPES measurement (typically much smaller than the total time-evolution) reducing the scaling from $O(N_t^3)$ to $O(N_t + \Tilde{N_t}^2)$, where $N_t$ is the total number of time steps and $\Tilde{N}_t$ scales with the width of the probe window with $\Tilde{N}_t \ll N_t$.  In the remainder of this perspective we will give a brief theoretical introduction to NEGFs before discussing the new method.  We will introduce the theory behind our scheme and provide several illustrative examples showing its ability to accurately capture dynamical correlations in time-resolved spectra, and it's scalability by showing simulation of a system several times larger than has previously been possible while including fully dynamical self-energy effects necessary to capture emergent phenomena.  We will conclude with an outlook of how the RT-DE can be used as the base of a framework for the \emph{ab-intio} simulation of time-resolved spectra and discussion of current and future work to make this a reality.

\section{Theoretical background}
\subsection{Notation}
In the remainder of this perspective we will make use of notation commonly used in the field of NEGFs.  The relevant notations  encountered in this work and their descriptions are provided in Table \ref{tab:notation}.
\begin{table}
    \centering
    \begin{tabular}{||c|c||}
    \hline
         Notation& Description  \\
         \hline
        & GF corresponding to\\$G^<(t,t')$ & occupied portion of the spectrum\\
        \hline
        & GF corresponding to\\$G^>(t,t')$ & unoccupied portion of the spectrum\\
        \hline
         & GF corresponding to \\$G^R(t,t')$&the full spectrum of the system\\
        \hline 
        & Self-energy computed with \\$\Sigma(t,t')$&many-body perturbation theory which \\ & can also have a R$/\lessgtr$ component\\
        \hline
        & GF computed \\$G^\mathrm{MF}(t,t')$&using a mean-field Hamiltonian\\
        \hline
        & Time and momentum resolved  \\$\mathcal{A}^{</\mathrm{R}}(k,\omega,t_p)$ & spectral function computed with \\&$G^<(t,t')$/$G^\mathrm{R}(t,t')$ centered around $t_p$\\
        \hline
    \end{tabular}
    \caption{Notation for the non-equilibrium \\Green's function formalism}
    \label{tab:notation}
\end{table}
\subsection{The time-resolved spectral function}

As mentioned in the introduction, one of the key experimental probes of quantum materials both in and out of equilibrium is ARPES and it's time-resolved extension\cite{Krausz_2009,Boschini_timeresolved_2024}.  Photoemission experiments probe the energy distribution of electrons in a material by using photons to eject electrons from a sample. By measuring the electron's outgoing kinetic energy, $E^\mathrm{kin}$, the electron's binding energy can be computed with $\varepsilon^\mathrm{binding} = E^\mathrm{kin} - \hbar\omega^\mathrm{photon}$, where $\hbar\omega^\mathrm{photon}$ is the energy of the photon. Equivalently however, $\varepsilon^\mathrm{binding}$ is the difference between the energy of the system before and after the removal of the electron. In other words $\varepsilon^\mathrm{binding} = E^N - E^{N-1}$, where $E^N$ is the energy of the original $N$ particle system and $E^{N-1}$ is the energy after removal of a specific electron.  This quantity is directly related to the GF, which is a probability amplitude of an electron (or hole) injected at one space-time coordinate to be found at a different space-time coordinate and is expressed mathematically below in terms of electron creation and annihilation operators, $c^\dagger(\mathbf{r},t)/c(\mathbf{r},t$, and the many-body ground-state wavefunction for the $N$ particle system, $|\Psi^N_0\rangle$.  We will now show that $\varepsilon^\mathrm{binding}$ can be extracted theoretically by considering the time evolution of the GF. 
\begin{align*}
    G(\mathbf{r},t; \mathbf{r}',t') &= \langle \Psi_0^N|c^\dagger (\mathbf{r},t)c(\mathbf{r}',t')|\Psi^N_0\rangle\\
        &= \langle \Psi_0^N|e^{-i\mathcal{H}t}c^\dagger(\mathbf{r}) e^{-i\mathcal{H}(t'-t)}c(\mathbf{r}')e^{i\mathcal{H}t'}|\Psi^N_0\rangle\\
        &=e^{-iE_0^N(t-t')}\langle \Psi_0^N|c^\dagger(\mathbf{r}) e^{-i\mathcal{H}(t'-t)}c(\mathbf{r}')|\Psi^N_0\rangle\\
        &=\sum_{i}e^{-iE_0^N(t-t')}\langle \Psi_0^N|c^\dagger(\mathbf{r}) e^{-i\mathcal{H}(t'-t)}|\Psi_i^{N-1}\rangle \langle \Psi_i^{N-1}|c(\mathbf{r}')|\Psi^N_0\rangle\\
        &=\sum_{i}e^{-iE_0^N(t-t')}\langle \Psi_0^N|c^\dagger(\mathbf{r}) e^{-iE_i^{N-1}(t'-t)}|\Psi_i^{N-1}\rangle \langle \Psi_i^{N-1}|c(\mathbf{r}')|\Psi^N_0\rangle\\
        &=\sum_{i}e^{-i(E_0^N-E_i^{N-1})(t-t')}\langle \Psi_0^N|c^\dagger(\mathbf{r})|\Psi_i^{N-1}\rangle \langle \Psi_i^{N-1}|c(\mathbf{r}')|\Psi^N_0\rangle\\
        &=\sum_i\psi^{N-1}_i(\mathbf{r})\psi_i^{N-1^*}(\mathbf{r}')e^{-i\epsilon_i^{\mathrm{binding}}(t-t')}.
\end{align*}
Here we have an included explicit coordinate $\mathbf{r}$, this can a spatial variable or thought of as a discreet index labeling spin, bands, orbitals etc,  $\varepsilon^\mathrm{binding}_i$ corresponds to the removal energy of the $i^{\mathrm{th}}$ electron and $\psi^{N-1}_i(\mathbf{r})$ is the corresponding Dyson orbital with coordinate $\mathbf{r}$.  In the above we have inserted a complete set of eigenstates and made use of the fact that there will be zero overlap between $c|\Psi^N_0\rangle$ and any state that differs by more than the removal of a single particle.
It is clear from the above expression that by Fourier transforming the GF to frequency space we gain access to the oscillation frequencies which correspond to the electron removal energies measured in photoemission experiments.  Expressed mathematically we have,
\begin{equation}\label{eq:eq_spectral_function}
    \mathcal{A}^{\mathrm{ARPES}}(\mathbf{k},\omega) = \int_{-\infty}^\infty dt \space e^{i\omega t}G(\mathbf{k},t),
\end{equation}
where we now include explicit momentum dependence in the spectrum by Fourier transforming the spatial variable of the GF. Experimentally this is achieved by also measuring the angle at which the electrons are ejected from which their momentum in the sample can be inferred.  We note that in using the equality between the left and right hand sides of equation \eqref{eq:eq_spectral_function} we have: 1) invoked the sudden approximation that assumes the electron is ejected instantaneously from the material and 2) neglected any effects related to the surface geometry of the material.  These additional approximations are commonly used in GF simulations\cite{martin2016interacting}. 

In non-equilibrium a pump-probe setup is used where the system is first excited from equilibrium and then probed with ARPES at different times during its time-evolution. The equivalent expression to equation \eqref{eq:eq_spectral_function} is more elaborate because: the system is no longer stationary and the signal will change depending on the probe delay relative to the pump and the probe duration and the GF depends explicitly on two-times instead of just a time difference.  The resulting expression is given below and the relevant details of its derivation can be found in Ref. \cite{freericks_2009},
\begin{equation}\label{eq:spectral_function_perspective}
        \mathcal{A}^{\mathrm{TR-ARPES}}(k,\omega,t_p) = \int dt dt'\space \mathrm{e}^{-i\omega(t-t')}\mathcal{S}(t-t_p)\mathcal{S}(t'-t_p) G(k,t,t')\\
\end{equation}
We now Fourier transform in the relative time coordinate of the GF, $t-t'$.  Furthermore, the functions $\mathcal{S}(t-t_p)$ are windowing functions that weight the Fourier transform around the time the system is probed ($t_p$).  Here we invoke the same additional approximations as we have for equation \eqref{eq:eq_spectral_function}, ie. the sudden approximation and neglecting surface effects.  Other formulations of the non-equilibrium spectral function exist, with the main one defined in terms of the the Wigner distribution function with the two-time GF as it's input function\cite{Stefanucci_2021,Freericks_2013}.  The expression given in equation \eqref{eq:spectral_function_perspective} however is the most physically motivated, taking into account the finite probing time of a real TR-ARPES measurement, and is the definition we use for calculations shown throughout this work
.
\subsection{Time evolving the non-equilibrium Green's function}
The prediction of the time-resolved spectral function requires an efficient and systematically improvable way of time-evolving the GF of a non-equilibrium system.  The GF formalism is naturally time-dependent and the equation of motion for the GFs in Table \ref{tab:notation} are given by
\begin{equation}\label{eq:GF_eom}
\begin{split}
\frac{\mathrm{d}G^{\mathrm{\lessgtr/\mathrm{R}}}(t,t')}{\mathrm{d}t} &= -ih^{(0)}(t) G^{\mathrm{\lessgtr/R}}(t,t') + I^{\mathrm{\lessgtr/R}}(t,t'),\\
I^{\lessgtr}(t_1,t_2) &= \int_{0}^{t_1} \mathrm{d}\bar{t} \Sigma^\mathrm{R}(t_1,\Bar{t})G^{\lessgtr}(\Bar{t},t_2) +\int_{0}^{t_2} \mathrm{d}\bar{t} \Sigma^{\lessgtr}(t_1,\Bar{t})G^\mathrm{A}(\Bar{t},t_2),\\
I^\mathrm{R}(t,t') &= \int_t^{t'}d\bar{t}\Sigma^{\mathrm{R}}(t,\bar{t})G^{\mathrm{R}}(\bar{t},t'),
    \end{split}
\end{equation}
and are referred to as the Kadanoff-Baym equations (KBEs).   Here $h^{(0)}(t)$ contains all one-body terms in the system's Hamiltonian, including coupling to external fields, and $\Sigma$ is the self-energy that encodes information about many-body interactions in the system.   This form of the KBEs assumes the state at $t=0$ has already been prepared in the correlated ground state of the system of interest.  The initial state can be generated through a thermal Matsubara integration along the imaginary time axis. We do not consider thermal effects in this manuscript but description of the full, temperature dependent KBEs can be found in Ref. \cite{stefanucci2013nonequilibrium}.    Neglecting temperature, a correlated initial state can be found by starting from a single particle ground state (non-interacting or mean-field) and adiabatically switching on the self-energy contribution to prepare the correlated ground state. Adiabatic switching poses a possible issue for initial state generation as it often fails in systems with strong correlations, however in practice the self-energy approximation, based on MBPT, also breaks down for strongly correlated systems.  Therefore, adiabatic switching is typically suitable in practical numerical simulations where the GF formalism is reliable.    We emphasize that equation \eqref{eq:GF_eom} is exact in principle but in practice the exact form of the self-energy operator, $\Sigma$, is not known and must be approximated with many-body perturbation theory\cite{Hedin_1965,Kadanoff_1962,martin2016interacting}.

\begin{figure}
    \centering
    \includegraphics[width=.5\linewidth]{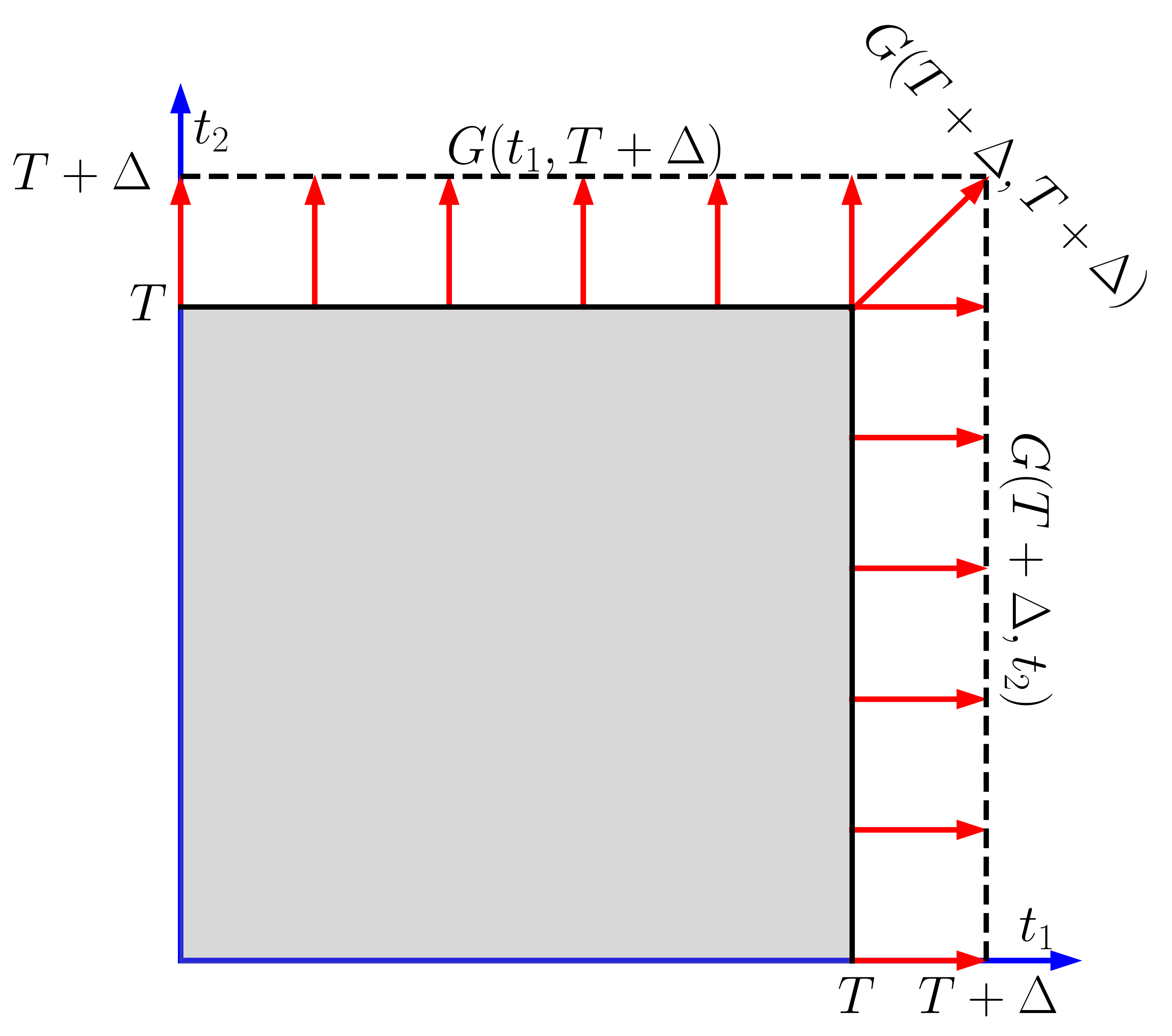}
    \caption{Schematic representation of the two-time grid on which the non-equilibrium Green's function lives.  Propagation of the NEGF requires time-evolution in both time variables at each time-step.}
    \label{fig:two_time_grid}
\end{figure}

The primary reason the NEGF formalism has not found a wide adoption, especially for ab-initio studies of materials, is due to the structure of the KBEs.  Unlike equilibrium, where time translation invariance dictates that the GF is a function of a time difference, the NEGF depends explicitly on two times and must be propagated simultaneously in both coordinates.  This is shown visually in Fig. \ref{fig:two_time_grid} where in order to time-step by one increment the GF must be propagated for all $t_1$ while keeping $t_2$ fixed and vice-versa, as well as along the time-diagonal.  Combined with the integral nature of the KBEs, the computational cost of these equations scales cubically with the number of time steps, making them prohibitively expensive in practical first principles calculations and they are extensively used only in application to various model systems\cite{Kwong_1998,Kwong_2000,Kremp_1999,Lake_1992,Schmitt_Rink_1988,Dahlen_2007,Banyai_1998,blommel_2025,Ong_2025}.

Overcoming the KBEs scaling issue has been one of the primary goals of NEGF research\cite{Lipavsky_1986,kaye_2021,Kaye_2023,schlunzen_2020,Perfetto_2022,chan_2021,stahl_2022,blommel_2024,zhu_2025,chan_2023}. In particular, in recent years this has given rise to several important developments towards practical \textit{ab-initio} simulations using NEGFs.  This includes practical approximations to the KBEs, the most widely used of these being the Hartree-Fock generalized Kadanoff-Baym ansatz (HF-GKBA)\cite{Lipavsky_1986} which is based on the factorization of the collision integral and neglects correlation effects coming from the non-equal time GF. The HF-GKBA has found great success in simulating the non-equilibrium density matrix even achieving linear scaling in the number of time-steps\cite{schlunzen_2020,Perfetto_2022,Pavlyukh_2022,Pavlyukh_2022_2,Pavlyukh_2022_3}.  Despite its widespread use and generally good performance, the HF-GKBA faces several significant limitations. First, it approximates the time-off-diagonal GF propagation at the Hartree-Fock level, neglecting dynamical correlation effects. As a result, it is suitable for describing the dynamics of the equal-time GF only. Second, the method inconsistently mixes dynamically correlated time evolution along the time diagonal with statically correlated evolution off of the time diagonal, introducing a potential source of error that is difficult to quantify. Finally, beyond modifying the self-energy, there is no clear or systematic framework for incorporating additional correlations into the off-diagonal components, hindering systematic improvement of the HF-GKBA.

Another direction that has been pursued is in finding systematic time compression via low-rank approximations to the full two-time GF. This leads to reduction in the computational cost of performing the integrations at each time-step of the time-evolution and can reduce the scaling to $O(N_t^2\log(N_t))$\cite{kaye_2021}.  In a similar vein, there has been investigation into truncating the temporal range of the collision integrals, which makes use of the fact that the influence of the systems past on it's present state will typically die off at long enough separations\cite{stahl_2022}.  This approach, while less controllable than the low-rank method reduces the scaling to linear in the number of time-steps.  Finally, a class of numerical approaches have been used to successfully reduce the scaling of the NEGF equations of motion and extrapolate time-evolutions based on some portion of the GF trajectory.  Most notable are 1) the use of recurrent neural networks (RNN)\cite{BASSI2024,zhu_2025} to learn the dependence of the collision integral on the current state of the system which reduces the IDEs to ordinary differential equations (ODEs) and 2) the use of dynamical mode decomposition (DMD)\cite{Yin_2022,Reeves_2023} to extrapolate dynamics based on initial time-evolutions, removing the need to perform explicit (expensive) time-evolution of the GF.  These approaches are particularly useful as they can be trivially adapted to work with various NEGF time-evolution schemes to offer further computational savings.  

Unfortunately in the context of first principles studies of time-resolved spectral properties the aforementioned schemes still leave room for improvement.  The HF-GKBA, despite being a state of the art method in the field of NEGFs, predicts time-resolved spectra that are effectively limited in quality by those produced by Hartree-Fock which include no dynamical correlations in TR-ARPES simulations\cite{joost_2017,Reeves_2024}. This is a result of the time-off-diagonal GF's time-evolution---a key ingredient in time-resolved spectra---being approximated at the Hartree-Fock level.  On the other hand the low-rank compression scheme, while retaining the same information as the full KBEs are still limited to relatively small model systems by the $O(N_t^2\log(N_t))$ scaling.  Despite their shortcomings for the desired goal of simulating time-resolved spectra of real experimental systems, the success of these approximation schemes in their own realms of applicability tells us a very important thing: the off-diagonal components of the NEGF and self-energy that appear in the KBEs are often amenable to some form of compression or approximation. Motivated by this observation and the clear need for an efficient first principles methodology for simulating time-resolved spectra we have recently developed a novel approach specifically designed to efficiently simulate (with $O(N_t)$ scaling) time-resolved spectra while systematically including dynamical correlations and many-body effects\cite{Reeves_2024}.  

\section{The Real-time Dyson Expansion}
In this section we discuss the real-time Dyson expansion (RT-DE)\cite{Reeves_2024}, that overcomes the scaling constraints of the KBEs and includes emergent features in time-resolved spectra. We will first give a theoretical overview of the RT-DE and discuss the steps involved in deriving the RT-DE equations of motion.  We will then proceed by analyzing the RT-DE to understand it's numerical scaling and the advantages it offers over existing NEGF approaches.  Finally we will demonstrate the RT-DE in several example systems: We provide benchmarks against spectra computed in exactly solvable systems, we compare to KBE results in Ref. \cite{Tuovinen_2020} for a system with an excitonic insulator ground state, and provide results for time-resolved spectra of a photo-excited 6-band semiconductor model.

\subsection{Dynamical self-energy corrections}
The self-energy is a functional of the full many-body GF and there exists a systematic procedure to generate successively improved self-energy approximations\cite{Hedin_1965,martin2016interacting,Mejuto_2022,Stefanucci_2014,Kadanoff_1962}.  The self-energy's dependence on the GF means the equations of motion require a self-consistent solution meaning that at each time-step $\Sigma$ is re-evaluated with an updated GF which is in turn used to update the GF and this is repeated until convergence.  This is the case for equilibrium as well as non-equilibrium problems.  However even in equilibrium this typically expensive self-consistency is often neglected for a lower cost and highly successful approach known as the one-shot self-energy correction.  This involves performing a reference calculation using a static mean-field method such as Hartree-Fock or density functional theory.  The associated single particle wavefunctions can then be used to create a static approximation to the GF, $G^\mathrm{MF}$.  Finally, the self-energy is constructed with with $G^\mathrm{MF}$ yielding a correction on top of the mean-field reference states. While this neglects the full self-consistent treatment such an approach offers consistent improvement over mean-field\cite{godby_1986,godby_1988,Chelikowsky1996QuantumTO,AULBUR2000,Klein_2023,rohlfing_2000,golze_2019}.

\begin{figure}
    \centering
\includegraphics[width=\linewidth]{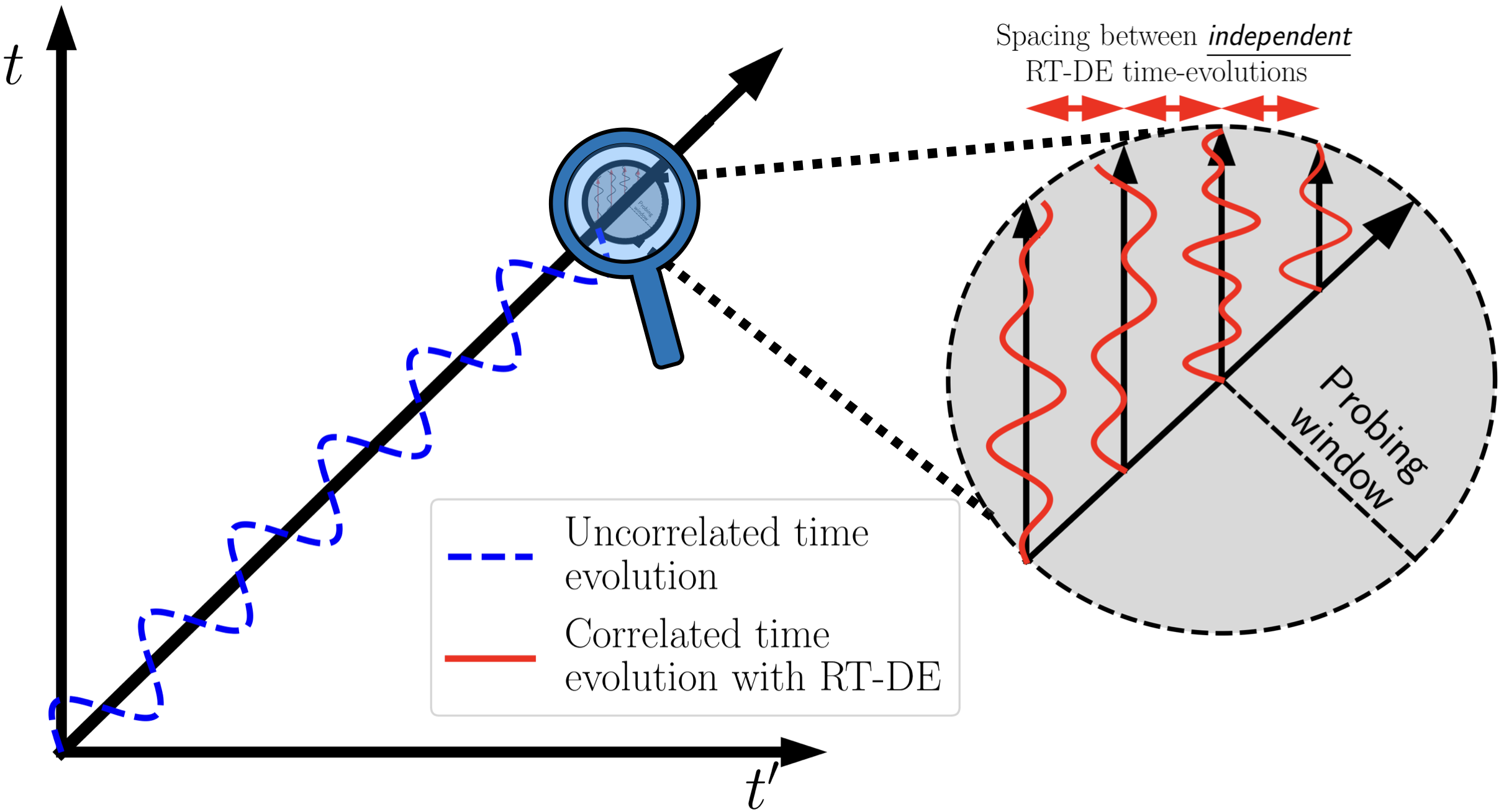}
    \caption{Schematic representation of the RT-DE procedure.  One first performs a time-evolution using an uncorrelated mean-field method.  Then within the probe window (typically much smaller than the full two-time grid) the RT-DE is employed to add dynamical correlations atop the mean-field reference state.  Each starting point along the time-diagonal can be run independently allowing for parallelization over temporal coordinates and the use of interpolation within the probe window (see Fig. \ref{fig:interpolation})}
    \label{fig:RTDE_schematic}
\end{figure}

Our RT-DE methodology extends this approach to the non-equilibrium regime by making one additional approximation to the KBEs (on top of the self-energy approximation).   We use a GF taken from a time-dependent static, ie. memory-less, GF simulation as our reference. Written mathematically we make the following  approximation,
\begin{equation}\label{eq:one_shot}
   \begin{split}
         \Sigma[G(t,t')] \approx \Sigma[G^{\mathrm{MF}}(t,t')],
   \end{split}
\end{equation} 
The idea behind the RT-DE is shown schematically in Fig. \ref{fig:RTDE_schematic}. One first performs a time evolution of the GF along the time-diagonal using a static mean-field Hamiltonian, a second time-evolution is then carried out for the non equal time GF. The key advantage is that time non-equal time GF components are reconstructed only within the probing region.  This second step adds dynamical corrections on top of the uncorrelated non-equilibrium reference state.  
Apart from removing the need to evaluate the self-energy self-consistently this approximation gives rise to another profound result that has the effect of reducing the scaling of the KBEs from cubic to linear in the total number of time-steps. This will be discussed in the next section as well as outlining the RT-DE equations of motion and the steps of its derivation.  Here we emphasize that since the RT-DE extends the equilibrium one-shot correction it should be straightforward to implement the RT-DE on top of real-time equilibrium GF codes. In particular we envisage combining the RT-DE with highly efficient real-time stochastic GF methods that have found great success in equilibrium\cite{vlcek_2017,vlcek_2018,vlcek_2019} to study large-scale non-equilibrium problems.

\subsection{Time-evolution of two-body propagators}
The key behind the reduced scaling of the RT-DE comes from the fact that the collision integrals from equation \eqref{eq:GF_eom}, that contain the contributions from the dynamical self-energy, are equivalently expressed in the following integral free form:
\begin{equation}\label{eq:coll_to_F}
    I^{\lessgtr/\mathrm{R}}_{im}(t,t') = -\sum_{klp}w_{iklp}\mathcal{L}^{C,\lessgtr/\mathrm{R}}_{lpmk}(t,t'),\\
\end{equation}
where we have explicitly included the matrix indices of the collision integral, $I^{\lessgtr/\mathrm{R}}(t,t')$, the two-body interaction tensor, $w_{ijkl}$, and an auxiliary two-particle GF, $\mathcal{L}^{C,\lessgtr/\mathrm{R}}_{lpmk}(t,t')$. $w_{ijkl}$ can be a statically screened or bare Coulomb kernel. $\mathcal{L}^{C,\lessgtr/\mathrm{R}}(t,t')$ holds the information of the self-energy and GF convolution and its exact form depends on the self-energy approximation used. It is the correlated portion of the full two-particle GF, meaning it cannot be expressed as a product of one-particle GFs, ie. $\mathcal{L}^C = \mathcal{L}^{(2)} - GG$ where $\mathcal{L}^{(2)}$ is the full two-particle GF.   This equivalence between $I^{\lessgtr/\mathrm{R}}(t,t')$ and $\mathcal{L}^{C,\lessgtr/\mathrm{R}}(t,t')$ comes from the derivation of the GF equation of motion resulting in the Martin-Schwinger hierarchy that links the time-evolution of an $N$ particle GF to that of an $N+1$ particle GF\cite{stefanucci2013nonequilibrium,martin_1959}. Clearly, knowing the time-evolution of $\mathcal{L}^{C,\lessgtr/\mathrm{R}}(t,t')$ removes the need to explicitly evaluate $I^{\lessgtr/\mathrm{R}}(t,t')$ and consequently the need to store the entire two-time GF.  The primary result of Ref. \cite{Reeves_2024} is to show that this can be achieved under the approximation outlined in the previous section by allowing for the derivation of a coupled set of ODEs for $G^{\lessgtr/\mathrm{R}}(t,t')$ and $\mathcal{L}^{C,\lessgtr/\mathrm{R}}(t,t')$.  In practice this approach is conceptually equivalent to ``upfolding'' the self-energy to a higher dimensional representation. As a result this step removes the time-non-local integration.

We will now outline the steps involved in deriving the RT-DE equations of motion.  For simplicity we derive the equations for the retarded GF [$G^\mathrm{R}(t,t')]$ using the second-Born self-energy, but emphasize that our scheme is generalizable to other components of the GF as well as other self-energy approximations including $GW$\cite{Reeves_2024}. For the second Born self-energy $\Sigma^{\mathrm{R}}$ is given by    
\begin{equation*}
    \begin{split}
        \Sigma^{\mathrm{R}}_{ij}[G(t,t')] &= -\sum_{klpqrs} w_{iklp}w_{qrsj}^\mathrm{x}\bigg{[}G^>_{lq}(t,t')G^>_{pr}(t,t')G^<_{sk}(t',t) 
        \\
        &\hspace{75mm}- G^<_{lq}(t,t')G^<_{pr}(t,t')G^>_{sk}(t',t)\bigg{]},\\
    \end{split}
\end{equation*}
where $w_{qrsj}^\mathrm{x} = w_{qrsj} - w_{qrjs}$ encodes the direct and exchange portion of the second-Born self-energy. Using the approximation of equation \eqref{eq:one_shot} we get,
\begin{equation*}
    \begin{split}
        \implies \Sigma^{\mathrm{R}}_{ij}[G^{,\mathrm{MF}}(t,t')] &= -\sum_{klpqrs} w_{iklp}w_{qrsj}^\mathrm{x}\bigg{[}G^{>,\mathrm{MF}}_{lq}(t,t')G^{>,\mathrm{MF}}_{pr}(t,t')G^{<,\mathrm{MF}}_{sk}(t',t) 
        \\&\hspace{60mm}- G^{<,\mathrm{MF}}_{lq}(t,t')G^{<,\mathrm{MF}}_{pr}(t,t')G^{>,\mathrm{MF}}_{sk}(t',t)\bigg{]}.\\
    \end{split}
\end{equation*}
From this and the definition of $I^{\mathrm{R}}(t,t')$ in equation \eqref{eq:GF_eom} we have
\begin{equation*}
\begin{split}
        I^{\mathrm{R}}_{im}(t,t') &= -\sum_{klp} w_{iklp}\bigg{[}\sum_{qrsj}\int_{t'}^t d\Bar{t}\medspace w_{qrsj}^\mathrm{x}\bigg{[}G^{>,\mathrm{MF}}_{lq}(t,\Bar{t})G^{>,\mathrm{MF}}_{pr}(t,\Bar{t})G^{<,\mathrm{MF}}_{sk}(\Bar{t},t) 
        \\&\hspace{60mm}- G^{<,\mathrm{MF}}_{lq}(t,\Bar{t})G^{<,\mathrm{MF}}_{pr}(t,\Bar{t})G^{>,\mathrm{MF}}_{sk}(\Bar{t},t)\bigg{]}G^{\mathrm{R}}_{jm}(\Bar{t},t')\bigg{]} \\
        &= -\sum_{klp}w_{iklp}\mathcal{L}^{C,\mathrm{R}}_{lpmk}(t,t').
\end{split}
\end{equation*}
Thus for the second-Born self-energy approximation $\mathcal{L}^{C,\mathrm{R}}(t,t')$ has the following form,
\begin{equation}\label{eq:explicit_F_form}
    \begin{split}
        \mathcal{L}^{C,\mathrm{R}}_{lpmk}(t,t') &= \int_{t'}^t d\Bar{t}\medspace w_{qrsj}^\mathrm{x}\bigg{[}G^{>,\mathrm{MF}}_{lq}(t,\Bar{t})G^{>,\mathrm{MF}}_{pr}(t,\Bar{t})G^{<,\mathrm{MF}}_{sk}(\Bar{t},t) 
        \\&\hspace{60mm}- G^{<,\mathrm{MF}}_{lq}(t,\Bar{t})G^{<,\mathrm{MF}}_{pr}(t,\Bar{t})G^{>,\mathrm{MF}}_{sk}(\Bar{t},t)\bigg{]}G^{\mathrm{R}}_{jm}(\Bar{t},t')\bigg{]}.
    \end{split}
\end{equation}
Taking the time derivative of the right hand side with respect to the integral's time argument (denoted as $\int$ in the subscript) gives,
\begin{equation*}
 \begin{split}
    \left[\frac{\mathrm{d}\mathcal{L}^{C,\mathrm{R}}_{lpmk}(t,t')}{\mathrm{d}t}\right]_{\int} &= \sum_{qrsj} w_{qrsj}^\mathrm{x}(t)\Big[ G^{>,\mathrm{MF}}_{lq}(t)G^{>,\mathrm{MF}}_{pr}(t)G^{<,\mathrm{MF}}_{sk}(t) - G^{<,\mathrm{MF}}_{lq}(t)G^{<,\mathrm{MF}}_{pr}(t)G^{>,\mathrm{MF}}_{sk}(t)\Big{]} G^{\mathrm{R}}_{jm}(t,t').
 \end{split}
\end{equation*}
Taking the derivative of the $t$ dependent mean-field GFs appearing in equation \eqref{eq:explicit_F_form} (denoted as $G^\mathrm{MF}$ in the subscript) and making use of the special form of the mean-field GF equations of motion, 
\begin{equation}\label{eq:MF_eom}
    \begin{split}
        \frac{\mathrm{d}G^{\lessgtr,\textrm{MF}}_{ij}(t,t')}{\mathrm{d}t} &= -i\sum_k h^{\mathrm{MF}}_{ik}(t)G^{\lessgtr,\textrm{MF}}_{kj}(t,t'),\\
         \frac{\mathrm{d}G^{\lessgtr,\textrm{MF}}_{ij}(t',t)}{\mathrm{d}t} &= i\sum_kG^{\lessgtr,\textrm{MF}}_{ik}(t',t)h^{\mathrm{MF}}_{kj}(t),\\
    \end{split}
\end{equation}
leads to the following
\begin{equation*}
    \begin{split}
&\left[\frac{\mathrm{d}\mathcal{L}^{C,\mathrm{R}}_{lpmk}(t,t')}{\mathrm{d}t}\right]_{G^\mathrm{MF}} = -i\sum_{x}\Big{[} h^\mathrm{MF}_{lx}(t) \mathcal{L}^{C,\mathrm{R}}_{xpmk}(t,t') + h^\mathrm{MF}_{px}(t) \mathcal{L}^{C,\mathrm{R}}_{lxmk} (t,t')-   \mathcal{L}^{C,\mathrm{R}}_{lpmx}(t,t') h^\mathrm{MF}_{xk}(t)\Big{]}.
    \end{split}
\end{equation*}
These are combined to give the RT-DE equations of motion for the second-Born self-energy,
\begin{equation}\label{RT-DE}
       \begin{split}
       \frac{\mathrm{d}G^{\mathrm{R}}(t,t')}{\mathrm{d}t} &= -i\left[h^{\mathrm{MF}}(t) G^{\mathrm{R}}(t,t') + I^{\mathrm{R}}(t,t')\right],\\
       I^{\mathrm{R}}_{im}(t,t') &= -\sum_{klp}w_{iklp}(t) \mathcal{L}^{C,\mathrm{R}}_{lpmk}(t,t'),\\
        \frac{\mathrm{d}\mathcal{L}^{C,\mathrm{R}}_{lpmk}(t,t')}{\mathrm{d}t} &=  \sum_{qrsj} w_{qrsj}^\mathrm{x}(t)\Big{[} G^{>,\mathrm{MF}}_{lq}(t)G^{>,\mathrm{MF}}_{pr}(t)G^{<,\mathrm{MF}}_{sk}(t) - G^{<,\mathrm{MF}}_{lq}(t)G^{<,\mathrm{MF}}_{pr}(t)G^{>,\mathrm{MF}}_{sk}(t)\Big{]} G^{\mathrm{R}}_{jm}(t,t') \\&\hspace{15mm}-i\sum_{x}\Big{[} h^\mathrm{MF}_{lx}(t) \mathcal{L}^{C,\mathrm{R}}_{xpmk}(t,t') + h^\mathrm{MF}_{px}(t) \mathcal{L}^{C,\mathrm{R}}_{lxmk} (t,t') - \mathcal{L}^{C,\mathrm{R}}_{lpmx}(t,t') h^\mathrm{MF}_{xk}(t)\Big{]}.\\
   \end{split}
\end{equation}
The RT-DE proceeds with an initial time evolution of $G^\mathrm{\lessgtr,MF}(t,t)$ using 
\begin{equation}\label{eq:diagonal_eom}
    \frac{dG^\mathrm{\lessgtr,MF}(t)}{dt} = -i [G^\mathrm{\lessgtr,MF}(t),h^{\mathrm{MF}}(t)],
\end{equation}
for some $h^{\textrm{MF}}(t)$. We emphasize this can be any static single particle Hamiltonian such that the time-evolution of $G^{\lessgtr,\textrm{MF}}$ follows the form of equations \eqref{eq:MF_eom} and \eqref{eq:diagonal_eom}.  Using this information for the time diagonal components in equation \eqref{RT-DE} we can time-step in the $t$ variable in the range $t'<t<T_{\mathrm{max}}$ with the following initial conditions for each $t'$,
\begin{equation*}
\begin{split}
        G^{\mathrm{R}}(t',t') &= -i,\\
        \mathcal{L}^{C,\mathrm{R}}(t',t') &= 0.
\end{split}
\end{equation*}
We refer the reader to Ref.\cite{Reeves_2024} for more detailed derivation and discussion, as well as derivations for the $GW$ self-energy approximation.  We note that a similar approach has also been applied to the time-diagonal GF in a linear scaling HF-GKBA scheme\cite{schlunzen_2020,joost_2020} but our work is the first application of this ODE approach for non-equal time GF and time-resolved spectral properties.  

\subsection{Scalability of the RT-DE}
Clearly, the trade-off for removing the explicit collision integrals is the need to now work with a four-index quantity, $\mathcal{L}^{C}_{ijkl}$.  This introduces two potential problems when considering scaling to large systems.  The first is related to the numerical scaling which scales with system size $N$ as $O(N^5)$ for second-Born and $O(N^6)$ for the $GW$ self-energy\cite{Reeves_2024}.  This scaling comes from tensor contractions appearing in equation \eqref{RT-DE}.  Fortunately, these types of tensor summations are highly amenable to parallelization on GPUs and CPUs.  Our current implementation of the RT-DE enables calculations to be run with GPU or CPU parallelization and allows for non-equilibrium spectra to be generated for multi-band systems several times larger than has previously been possible while accounting for dynamical correlations.

Another avenue for optimization is through compression of the four-index tensor $\mathcal{L}^{C,\lessgtr/\mathrm{R}}$.  It is likely the case, especially as we increase the size of the system, that $\mathcal{L}^{C,\lessgtr/\mathrm{R}}$ will have some degree of compressibility through a low-rank approximation.  This can be applied to $\mathcal{L}^{C,\lessgtr/\mathrm{R}}$ and the summations recast in a way that their scaling depends more favorably on the system size as well as the rank of $\mathcal{L}^{C,\lessgtr/\mathrm{R}}$.  This includes approaches such as Cholesky decomposition and tensor cross interpolation\cite{Pedersen_2024,Fernandez_2022,OSELEDETS2010}. The implementation of tensor decomposition in the RT-DE is a current focus of ours and combining this with our already GPU parallelized code will allow us to push to systems orders of magnitude larger than what has previously been possible.  

A final route to scalability of the RT-DE, that has already been briefly mentioned, is the leveraging of stochastic GF algorithms.  These have been used to simulate systems with tens of thousands of electrons and have the advantage of already being used in real-time (as opposed to frequency space common to many equilibrium GF methods)\cite{vlcek_2017,vlcek_2018,vlcek_2019}.  The combining of the RT-DE with these stochastic real-time methods is a theoretically straightforward extension that would allow for the first-principles simulation of non-equilibrium quantum systems with thousands of electrons.  

The second potential issue introduced by time-evolving the two-particle GF $\mathcal{L}^{C}$ is in the memory storage requirements. $\mathcal{L}^C(t,t')$ now contains $N^4$ elements compared to $N^2$ for the single particle GF, where $N$ is the systems size.  At first it may seem that storing this four index quantity, even just within the probing window of the full two-time grid, would lead to prohibitively expensive memory requirements for systems beyond $N\sim O(10)$. However, this is not the case in practice:  Firstly, we propagate ODEs instead of IDEs, so that each time-evolution in the $t$ direction only requires storage of $\mathcal{L}^C$ at the \emph{current} time-step.  Secondly the equations of motion can be run independently for each starting $t=t'$, meaning that even for systems large enough that running each propagation simultaneously is an issue --- scaling as $\tilde{N}_t\times N^4$ for $\tilde{N}_t$ RT-DE starting points --- the problem can be broken down into smaller components and run sequentially requiring at most the storage of a single $\mathcal{L}^C$. Ultimately in order to extract single particle spectral properties the GF will be needed and so the memory requirements in reality will scale as $\tilde{N}_t^2\times N^2$ compared to the $N_t^{2}\times N^2$ of the KBEs where $\tilde{N}_t \ll N_t$. Thus in addition to saving computational cost by reducing scaling from cubic to linear in the number of time-steps the RT-DE also offers huge savings in memory, necessary for simulating large scale systems.  

We highlight the importance of the ability to parallelize over temporal coordinates as well as other degrees of freedom. This adds even further to the scalability of the RT-DE and is relatively simple to implement simply requiring the starting points within probe region to be sub-divided evenly depending on the desired number of independent runs.  Furthermore it can be trivially combined with the more traditional parallelization schemes discussed above.  Finally, as we will discuss and demonstrate later, this independence of each RT-DE time-evolution can be combined with interpolation in the time-domain to reduce the number of RT-DE time-evolutions that must be performed along the $t$ direction.

\section{Applications of the RT-DE}
\subsection{Finite system benchmarks}
\begin{figure}
    \centering
    \includegraphics[width=\linewidth]{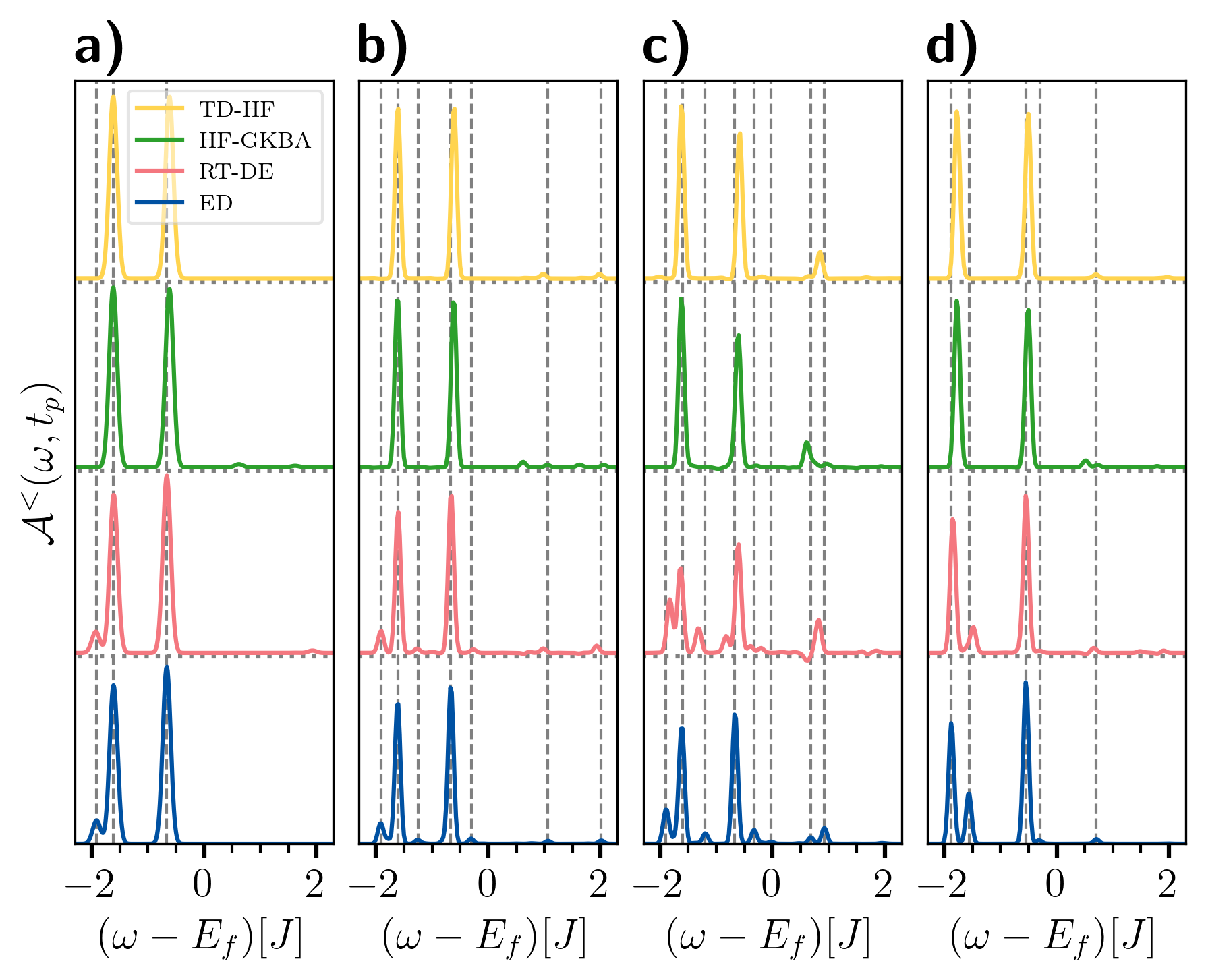}
    \caption{Exact diagonalization benchmarks for the RT-DE (pink) and HF-GKBA (green) with the second-Born self-energy and time-dependent Hartree-Fock (yellow).  Results show for occupied ground and excited state spectra for the four-site Hubbard model with $U=1.0J$.  Computed for the model in equation \eqref{eq:MB_ham} with the following parameters: $U=1.0J$,  $N_s = 4$, $t_0 = 5J^{-1}$ $T_p = 0.5J^{-1}$, $N_q=2$ and $T_{\mathrm{max}} = 200J^{-1}$. a) $h^{\mathrm{N.E}} = 0$, b) $h^{\mathrm{N.E}} = h^{\textrm{Short}}$ and $E=0.5J$, c) $h^{\mathrm{N.E}} = h^{\textrm{Long}}$ and $E=0.5J$, d) $h^{\mathrm{N.E}} = h^{\textrm{Quench}}$ and $E=1.0J$.  Reprinted with permission from Ref. \cite{Reeves_2024}. Copyright 2024 American Physical Society. }
    \label{fig:finite_system}
\end{figure}
We will first present several benchmarks of the RT-DE for a finite system accessible to exact diagonalization\footnote{The results in this section are reproduced from Ref. \cite{Reeves_2024} with permission of the publisher}.  We use the Hubbard model Hamiltonian written explicitly in second quantized form below  
\begin{equation}\label{eq:MB_ham}
\begin{split}
        \mathcal{H}  &= -J\sum_{\langle i,j\rangle}c^\dagger_ic_{j} + U\sum_{i}n_{i\uparrow}n_{i\downarrow} + \sum_{ij}h_{ij}^{\mathrm{N.E}}(t)c_{i}^\dagger c_j.\\
\end{split}
\end{equation} 
This Hamiltonian describes a lattice of electrons with nearest neighbor hopping, determined by $J$, and an interaction between electrons of opposite spin on the same lattice site, determined by $U$. The simulations are done with $U=J=1$ and $N_s=4$. The system is prepared in its ground state and excited with different non-equilibrium Hamiltonians which are given by
\begin{equation}\label{eq:h_NE}
     h_{ij}^{\mathrm{N.E}}(t) = \begin{cases}
       h^{\textrm{short}}_{ij}(t)&=\delta_{ij}E\cos\left(\frac{\pi r_i}{2}\right)\exp\left({-\frac{(t-t_0)^2}{2T_p^2}}\right)\\
       h^{\textrm{long}}_{ij}(t)&= \delta_{ij}Er_i\exp\left({-\frac{(t-t_0)^2}{2T_p^2}}\right)\\
       h^{\textrm{quench}}_{ij}(t)&= \delta_{ij}E\Theta(t)\Theta(N_q-i-1) 
    \end{cases}
\end{equation}
Here $r_i = \frac{N_s-1}{2} -i$ and $\Theta(t)$ is the Heaviside step function. 

The results shown in Fig. \ref{fig:finite_system} are computed with equation \eqref{eq:spectral_function_perspective} using $G^<(t,t')$ and correspond to the occupied portion of the spectrum.  Fig. \ref{fig:finite_system} a) shows the ground state and Fig. \ref{fig:finite_system} b)-d) show the spectra after excitation by the Hamiltonians given in equation \eqref{eq:h_NE}.  Detailed analysis and discussion of these results can be found in Ref. \cite{Reeves_2024}.  Here we highlight two important points.  Firstly, we can see the inability of the HF-GKBA to produce results beyond those given by TD-HF, which shows the need for a method beyond the HF-GKBA (a current state of the art method for NEGF propagation).  Second and most importantly the results demonstrate the ability of the RT-DE to quantitatively capture ground and excited state spectral features that are missed by the HF-GKBA and TD-HF results and are the result of dynamical correlations in the system. 

\subsection{Excitonic insulator}
The next application of the RT-DE we discuss is for the prediction of ground state properties of a model excitonic insulator\cite{Jerome_1967,Kaneko_2025}, a type of correlated insulator that host excitons (bound electron-hole pairs) in its ground state.  This system can be thought of as analogous to a superconducting system where Cooper pairs (bound pairs of electrons) are replaced by excitons and the superconducting pairing is replaced by Coulombic electron-hole attraction.  

\begin{figure}
    \centering
    \includegraphics[width=.5\linewidth]{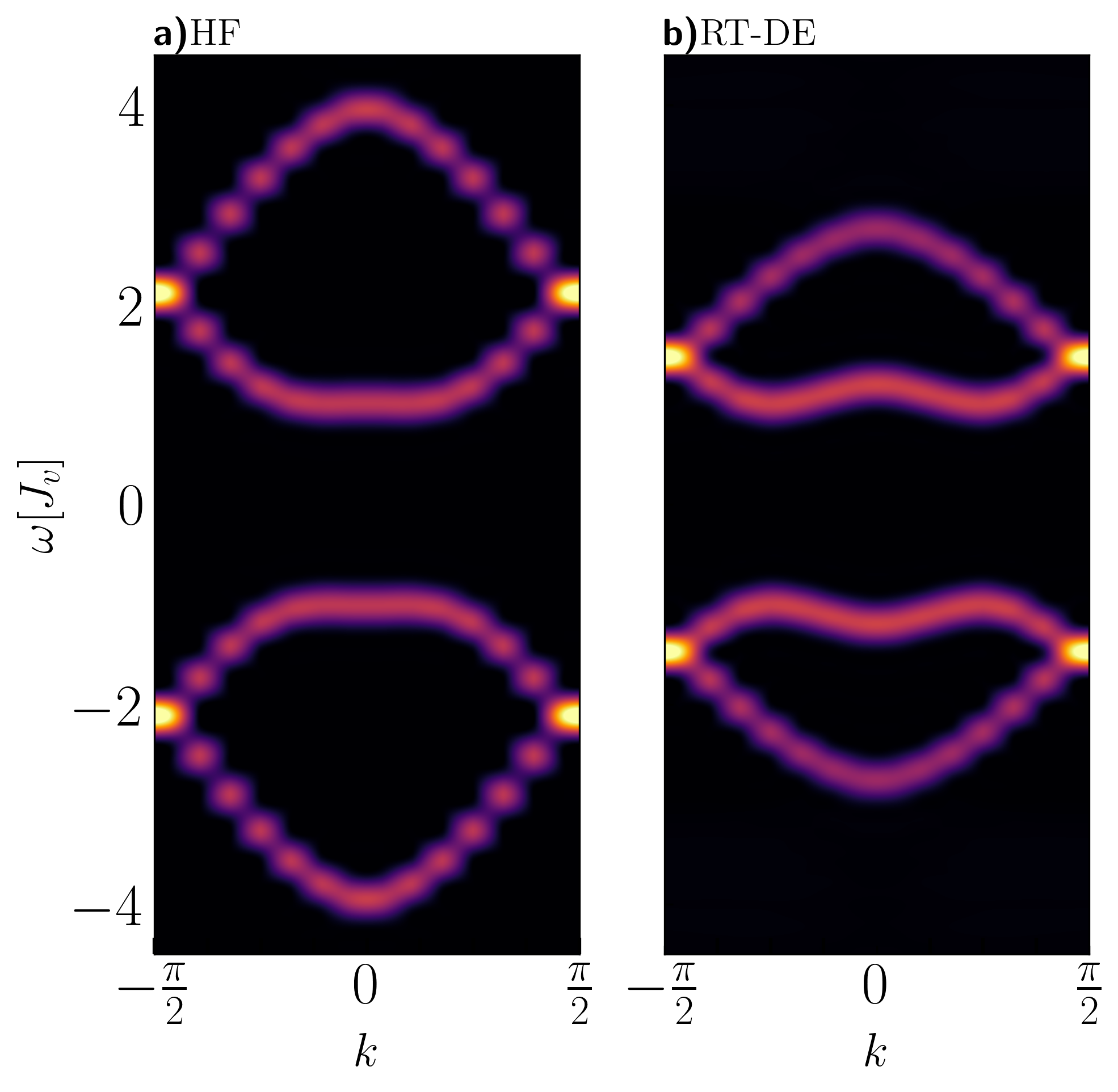}
    \caption{Ground state spectrum for excitonic insulator described by equation \eqref{eq:excitonic_ham} with $\Delta = 1.4J$ and $U=3.5J$.  Result computed with 28 sites backfolded to give $N_k=14$.  The result is computed using a) HF (equivalent to $\mathcal{L}^C =0$ in the RT-DE equations of motion) and b) the RT-DE evaluated with the second-Born self-energy}
    \label{fig:excitonic_insulator}
\end{figure}

Fig. \ref{fig:excitonic_insulator} shows the full (occupied and unoccupied) ground state spectrum computed using $G^\mathrm{R}(t,t')$ computed with Hartree-Fock and the RT-DE using the second-Born self-energy for the Hamiltonian given below.
\begin{equation}\label{eq:excitonic_ham}
    \mathcal{H}  = -J\sum_{\langle i,j\rangle,\alpha}c^\dagger_{i,\alpha}c_{j,\alpha} + \Delta\sum_{\langle i,j\rangle}n_{i,c} + U\sum_{i}n_{i,c}n_{i,v}
\end{equation}
This is a two-band model of spinless fermions on a periodic lattice where $c$ and $v$ correspond to the conduction and valence band respectively.  Similarly to the previous model the $J$ determines the strength of hopping between nearest neighbor sites in a given band and $U$ determines the interaction strength between particles on the same site and different bands. $\Delta$ is the onsite potential for each orbital.  Despite the translational invariance---under translation by a single site index---of the full many-body Hamiltonian, the excitonic insulating states can break this symmetry through the phenomenon of spontaneous symmetry breaking\cite{Beekman_2019}.  To preserve periodicity it is necessary to double the unit cell, making it pairs of neighboring sites. This doubling of the unit cell results in the Brillouin zone being reduced and ``back-folded'' into $k\in [-\frac{\pi}{2},\frac{\pi}{2}]$.  The symmetry breaking is taken into account as outlined in Ref.\cite{Tuovinen_2020} and we use the parameters from this same reference for direct comparison between the RT-DE and KBE spectra: $J=1, \Delta = 1.4J$ and $U=3.5$.  

Two striking differences exist between the mean-field (HF) and dynamically correlated (RT-DE) method.  First is the renormalization of the band-width which is much narrower in the RT-DE calculation compared to HF and which we attribute to non-local self-energy effects.  Second, the RT-DE result shows a pronounced back-bending in the dispersion around the gamma point, consistent with an excitonic ground state\cite{Tuovinen_2020,Shao_2024,Werdehausen_2018,Huang_2024}.  In essence, this is a signature of a large population of interacting excitons that comes from observations in superconducting systems\cite{Rinott_2017,Takahiro_2020} where analogy is drawn between Cooper pairs and excitons.   As RT-DE includes explicit propagation of a two-particle GF ($\mathcal{L}^C$) it captures the formation of excitons, an emergent feature missing in the mean-field result. Both of these features are qualitatively and quantitatively consistent with what is predicted in Ref. \cite{Tuovinen_2020} using the full KBE treatment.  This example demonstrates the RT-DEs capability to predict properties of systems with non-trivial correlated ground states.  In the future the RT-DE will allow us to perform time-resolved simulations of driven excitonic systems and to determine spectral signatures of excitonic phenomena such as melting of excitonic order or the excitonic BCS-BEC crossover\cite{Bretscher2021,Takamura_2024,Perfetto_2019,Phan_2010}.  Of particular interest is the comparison of the RT-DE and KBEs for time-resolved properties driven excitonic insulators, though KBE propagation's are limited to relatively small systems and short simulation times.   

\subsection{Scaling up the RT-DE for multiband systems}

\begin{figure}
    \centering
    \includegraphics[width=\linewidth]{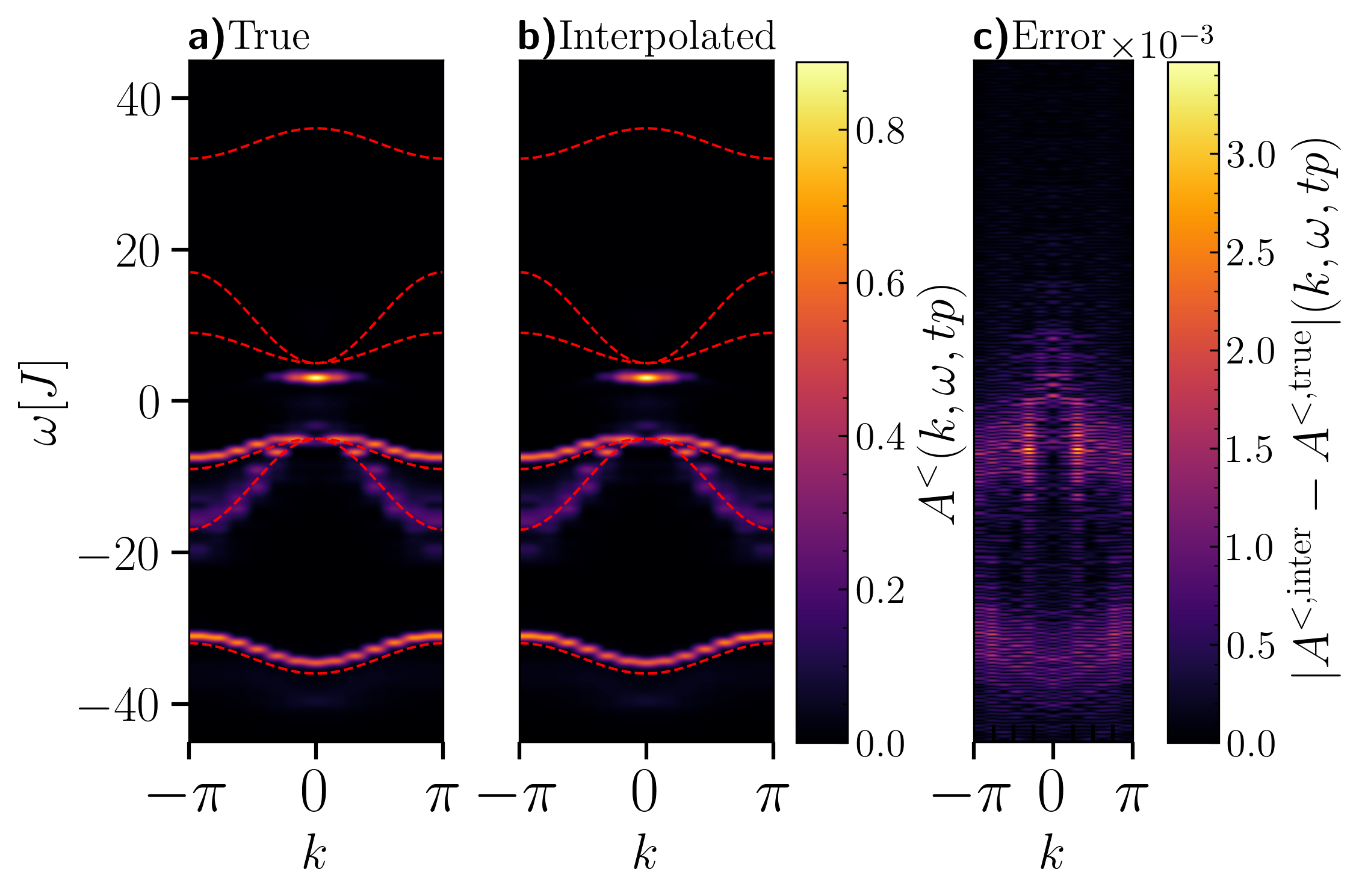}
    \caption{a) Time-resolved spectral function of the system described by equation \eqref{eq:six_band} after photo-excitation by pumping pulse resonant with the fundamental band-gap.  The following parameters are used in the model: $J_\alpha = \{1,-1,-3,3,1,1\}$,   $U = 4J$, $E=0.25J$, $w_p = 8.0J$, $t_0 = 10J^{-1}$, $T_p = 10J^{-1}$ and $\varepsilon = 2$. b) Spectral function computed using coarsened grid of RT-DE starting points interpolated with cubic interpolation.  Original data uses $dt=0.025$ between starting points on the time-diagonal for the RT-DE propagation. Using our interpolation scheme it is possible to coarsen this by as much as 40 times with minimal loss of accuracy.  Interpolated data reconstructed using $dt=1.0$ between starting points, while RT-DE propagation still uses $dt=0.025$.  c) shows the error introduced by the interpolation scheme which is negligible even for this relatively coarse grid.  We note that the interpolation is done parallel to the time-diagonal as opposed to perpendicular/parallel to the $t$ axis.}
    \label{fig:interpolation}
\end{figure}

Optimization of the RT-DE for use on GPUs has allowed for simulation of multiband systems and further work combining GPU parallelization with the CSPACER software library\cite{Ibrahim_2020} has opened the door to simulation of systems far larger than has previously been possible with the KBEs or even the HF-GKBA,  both of which are typically applied to one or two-band systems with a few tens of sites/$k$-points or to small molecular/atomic systems \cite{Kwong_1998,Banyai_1998,Dahlen_2007,perfetto_2019_2,Perfetto_2022,Bonitz_2013,Tuovinen_2020,Schuler_2020,Hermanns_2014,Schlunzen_2017}.  To illustrate the RT-DE's capabilities we show results for a six band model obtained with regular NERSC computing facilities.  In Fig. \ref{fig:interpolation} we show the spectrum of a six band Hamiltonian optically excited by an external field. 
\begin{equation}\label{eq:six_band}
    \begin{split}
        \mathcal{H} &= \sum_{\alpha,\beta \atop{i,j,\sigma}}h^{\alpha\beta}_{ij}(t)c^{\dagger\alpha}_{i,\sigma}c^{\beta}_{j,\sigma}  + U\sum_{i,\alpha\neq\beta}n_{i\alpha}n_{i\beta} + U\sum_{i,\alpha}n_{i\alpha\uparrow}n_{i\alpha\downarrow}+ U\sum_{i\neq j,\alpha,\beta} \frac{n_{i\alpha}n_{j\beta}}{\varepsilon|i-j|}\\
        h^{\alpha\beta}_{ij}(t) &= \underbrace{J_\alpha\delta_{\alpha\beta}\delta_{\langle i,j\rangle}+ \epsilon_\alpha\delta_{\alpha\beta} }_{h^{(0)}}+ \underbrace{\delta_{ij}(1-\delta_{\alpha\beta})
        E\cos(\omega_p(t-t_0)) \mathrm{e}^{-\frac{(t-t_0)^2}{2T_p^2}}}_{h^{\mathrm{drive}}(t)}.
    \end{split}
\end{equation}
The Hamiltonian describes a periodic lattice with orbital dependent hopping between neighboring sites (determined by $J_\alpha$) and long-range density-density interactions between electrons on different sites and in different orbitals (determined by $U$ and $\varepsilon$).   Here we have chosen the following parameters: $J_\alpha = \{1,-1,-3,3,1,1\}$,   $U = 4J$, $E=0.25J$, $w_p = 8.0J$, $t_0 = 10J^{-1}$, $T_p = 10J^{-1}$ and $\varepsilon = 2$.  The onsite potentials are fixed by enforcing, a gap of $E_g=15J$ between the edges of bands one and three, a fundamental gap of $E_{\textrm{fundamental}} = 10J$, and a gamma point degeneracy between bands two and three.  Furthermore the model is symmetric around the fermi energy, $E_\mathrm{F} = 0$.

The red dashed lines in Fig. \ref{fig:interpolation} show the equilibrium band positions and we plot the occupied portion of the spectrum only to show the photo-excitation clearly.  The model exhibits renormalization and band-broadening effects consistent with previous observations in other systems\cite{Reeves_2024,Reeves_2025_2}.  

This model primarily serves as a demonstration of what can be achieved using the RT-DE formalism as benchmarking in a system of this size is effectively impossible, even with an approximate method such as the KBEs.  One route forward is comparing to Bethe-Salpeter calculations in a linear response regime (ie. in the low photodoping limit) but benchmarking beyond linear response will be a challenge due to lack of reference methods.  In other words RT-DE allows for simulation of strongly driven (out of equilibrium) systems which cannot be described by other methods.  However, the RT-DE does give a feasible way to simulate experimentally relevant systems which can be a more direct point of comparison.  This motivates our current and future work of performing first principles simulations using the RT-DE.

This result also demonstrates the use of the RT-DE interpolation scheme.  Fig. \ref{fig:interpolation} b) shows a spectral function reproduced by performing the RT-DE time-evolution from starting points along the time diagonal with a 40x increased spacing (coarser time-grid).  The RT-DE propagation in the time-off-diagonal direction is still performed with the original time-step.  Using this coarsened grid the Green's on the dense time-grid is reconstructed with a cubic spline interpolation.  Since we coarsen the starting points by a factor of 40 this translates into a forty-fold savings in the RT-DE calculation with a difference in the reconstructed spectral function of only around $10^{-3}$, which is negligible .  Thus, even this simple form of interpolation can add to the savings offered through the RT-DE. Potential improvements to our interpolation scheme will be discussed in section \ref{sec:summ_and_outlook}

In the final section we  will provide an outlook on how the RT-DE can serve as the core in a framework, that combined with further optimizations and practical numerical techniques, can be used for the simulation of large-scale realistic systems.

\section{Summary and Outlook}\label{sec:summ_and_outlook}

In this perspective we have discussed the need for an efficient framework for the simulation of time-resolved spectral properties of quantum many-body systems.  We have introduced the RT-DE as a technique that addresses this need.  Based on the simultaneous propagation of one and two-particle GFs in replacement of solving IDEs for the single particle GF alone, it overcomes scaling limitations present in the NEGF formalism without neglecting dynamical effects important for a range of non-equilibrium phenomena such as properties of excitonic systems and band-gap renormalization\cite{Reeves_2024,Reeves_2025_2}. We have discussed routes for continuing to scale the RT-DE to larger and larger systems. Some of these, such as CPU/GPU/time-grid parallelization and interpolation have already been implemented while some such as tensor decompositions of $\mathcal{L}^C$ will be implemented in future versions of the RT-DE. We concluded this perspective with several demonstrations of the RT-DE's capabilities to capture spectral features beyond the mean-field level and it's scalability to large systems compared to what has been studied previously.  We believe the RT-DE holds high potential as the basis of a framework for the efficient simulation of non-equilibrium quantum systems. We hope that this perspective shows the merits of our method and leads to broad interest in use, future development and improvement of the RT-DE.

We will finish this manuscript with a brief outlook on several ongoing and future development plans for the RT-DE.

\begin{enumerate}[(i)]
    \item \textbf{Cost-saving numerical techniques:}
    Despite the savings offered by the RT-DEs scaling in the number of time-steps the calculations can still have high cost per time-step.  Numerical extrapolation techniques such as the previously mentioned DMD\cite{Yin_2022,Reeves_2023} can be included within the RT-DE framework as a post-processing step to reduce the number of time-steps that must be explicitly evaluated.  The amount of explicit time-stepping can be further reduced by leveraging the interpolation that has already been demonstrated in Fig. \ref{fig:interpolation}.  Future work will improve upon this relatively simple fixed spacing interpolative scheme by allowing for adaptive interpolation\cite{blommel_2024}. This will minimize the number of RT-DE update steps without sacrificing accuracy and remove the need to laboriously test many different interpolation spacings for each different problem.  Machine learning in the form of recurrent neural networks (RNN) has also been used to reduce the cost of NEGF propagation\cite{BASSI2024,zhu_2025}.  We are currently investigating how this can be integrated with the RT-DE time-evolution as an advanced extrapolation \emph{or} interpolation tool.  
    \item \textbf{Advanced self-energy approximations:}  Currently our results make use of the second Born self-energy.  This is commonly used in NEGF calculations due to its low cost of implementation and relatively good accuracy.  However, the practical workhorse of GF calculations in equilibrium \emph{ab-initio} settings is the $GW$ approximation, thus it is desirable to work with this also in non-equilibrium.   We have shown explicitly that an RT-DE equation of motion can be derived for the $GW$ approximation\cite{Reeves_2024} and our theory can be extended to T-matrix and the dynamically screened ladder approximation\cite{joost_2022,joost_2020}.  The use of these approximations is rare with the KBEs\cite{Schuler_2020,Schlunzen_2016,Schlunzen_2016_2,Schlunzen_2017} due to the high cost on top of the KBEs poor scaling, but the savings offered by the RT-DE will allow for use of these more advanced self-energy approximations in practical calculations.  
    \item \textbf{Tensor compression:} Tensor contractions appear throughout physics and algorithms for reducing their scaling\cite{Evenbly_2022,Ran_2020,Khoromskij_2018}, typically based on decomposition into sums of tensors of lower order, are plentiful.  These decompositions can reduce the cost of tensor contractions, typically having lower scaling with size of the tensors and gaining a dependence on the rank or approximate rank of the full tensor.  Implementing this of course adds additional steps and cost for the decomposition steps, however this problem can take advantage of existing, highly optimized linear algebra software libraries to perform these steps efficiently.  This optimization should be relatively straightforward using an existing algorithm such as tensor cross interpolation, but will require further investigation into the rank structure of $\mathcal{L}^C$. 
    \item \textbf{Extending stochastic real-time GF methods to non-equilibrium:} The final future direction we will mention is the extension of our groups existing real-time stochastic codes from equilibrium to non-equilibrium systems using the RT-DE formalism.  This represents the most technically challenging of these points but it also holds the most potential for the simulation of large scale systems from first principles.  The stochastic GF approach already uses a real-time formulation and so the extension is conceptually straightforward. The scaling of the approach allows for simulation of systems with thousands of electrons and could be used to reduce the scaling even further than the tensor compression approach, especially when considering very large extended systems.  
\end{enumerate}

\section{Acknowledgements}
This material is based upon work supported by the U.S. Department of Energy, Office of Science, Office of Advanced Scientific Computing Research and Office of Basic Energy Sciences, Scientific Discovery through Advanced Computing (SciDAC) program under Award Number DE-SC0022198. This research used resources of the National Energy Research Scientific Computing Center, a DOE Office of Science User Facility supported by the Office of Science of the U.S. Department of Energy under Contract No. DE-AC02-05CH11231 using NERSC Award No. BES-ERCAP0032056.  Reeves is supported by the National Science Foundation through Enabling Quantum Leap: Convergent Accelerated Discovery Foundries for Quantum Materials Science, Engineering and Information (Q-AMASE-i) award number
DMR-1906325. 

\bibliography{Bib}

\end{document}